\documentstyle[aps,prl,epsfig]{revtex}

\tightenlines

\newcommand{\be}{\begin{equation}}
\newcommand{\ee}{\end{equation}}

\newcommand{\bmath}{\begin{mathletters}}
\newcommand{\emath}{\end{mathletters}}

\begin{document}

\title{\Large{\bf Magnetic ordering of itinerant systems; the role of kinetic exchange interaction}}

\vskip0.5cm 

\author{ G. G\'{o}rski and J. Mizia }

\address{Institute of Physics, University of Rzesz\'{o}w, ulica Rejtana 16A, \\
35-958 Rzesz\'{o}w, Poland\\}

\maketitle

\vskip0.5cm 


\begin{abstract}

\noindent
The possibility of ferromagnetic ordering is revisited in the band model. The coherent potential approximation decoupling has been used for the strong on-site Coulomb 
interaction. The driving forces towards the ferromagnetism are the on-site and inter-site molecular fields coming from different Coulomb interactions. Another driving force is the 
lowering of the kinetic energy with growing magnetic moment coming from the dependence of the hopping integrals on occupation of the neighboring sites involved in hopping. 
This effect is described by the hopping interaction, $\Delta t$ , and by what we call the exchange-hopping interaction, $t_{ex}$. The exchange-hopping interaction, which is the 
difference in hopping integrals for different occupation of neighboring lattice sites, acts in analogous way to the Hund's magnetic exchange interaction. The results are calculated 
for semi-elliptic density of states (DOS) and for the distorted semi-elliptic DOS with the maximum around the Fermi energy. They show a natural tendency towards the magnetic 
ordering at the end of the 3d row for the DOS with maximum density around the Fermi energy, when the hopping integrals grow with the occupation of the neighboring lattice 
sites. 

\end{abstract}

\vskip0.5cm



\vskip2.0cm 

\noindent {\Large {\bf 1. Introduction
} }

\vskip0.5cm

The basic model for magnetic ordering of itinerant electrons in solids is the Hubbard model \cite{1}. The largest interaction in this model is the on-site Coulomb repulsion, 
$U=\left({i,i|1/r|i,i}\right)$ , where $i$ is the lattice site index. In the mean-field approximation, the Hubbard model leads to the well-known Stoner model for magnetism \cite{2}. The 
Coulomb constant $U$ coming out of the Stoner condition for creating ferromagnetism is large, i.e. of the order of bandwidth. On one hand it can be justified by the existing strong 
Coulomb interaction, but on the other hand for such a strong interaction, one can not use mean-field approximation.

This has prompted attempts to treat the problem within a many body theory. The most significant approach, which we will mention in here, is the Hubbard I and conventional CPA 
\cite{3}. Unfortunately, like many other approaches, they also failed to bring any type of ferromagnetic ordering \cite{4}. They did not produce the spin-dependent band shift 
necessary for a ferromagnetic ordering which is why the new versions of the conventional CPA are still being created.

This is why in our previous paper \cite{5} we investigated the itinerant model for ferromagnetism with both single-site and two-site electron correlations. We included also the 
band degeneration into the model, what has allowed us to consider the on-site exchange interactions in the Hamiltonian. The modified Hartree-Fock approximation for the two-site 
interactions was used, which gave us the relative spin band broadening of one spin band with respect to the other. This was in addition to the shift in position of majority and 
minority spin bands. Despite the use of a traditional CPA approach the qualitatively new results were obtained which brought the constant of the mean-field creating magnetism 
almost to zero.

Quite recently some authors (see Hirsch's, Ref. \cite{6}) have pointed out that in the transition to the ordered state not only the potential energy is decreased but also the kinetic 
energy may be lowered. This lowering of the kinetic energy is something new in magnetism since the whole previous development has assumed that the kinetic energy grows 
during the transition to magnetic state, as oppose to the decrease in potential energy. The balance of these two energies resulted in the Stoner criterion for ferromagnetism.

The decrease of the kinetic energy, in this new approach, comes from the dependence of hopping integrals on occupation of the neighboring lattice site involved in hopping. It is 
described in our paper by two inter-site interactions, which are the hopping interaction, $\Delta t$, and the new interaction, which we call the exchange-hopping interaction, 
$t_{ex}$. The hopping interaction is defined as  $\Delta t=t_0-t_1$, where $t_0$ is the hopping integral when no other electron is present on sites involved in hopping and $t_1$ 
is the integral for hopping in the presence of one electron with the opposite spin on any of the two sites. The exchange-hopping interaction it is the difference of hopping 
integrals for different occupations of neighboring lattice sites given by Eq. (\ref{5}) below. For some parameters this interaction is decreasing the kinetic energy of the system 
during transition from paramagnetic to the ordered ferromagnetic state. Both these interactions are treated in the linear Hartree-Fock approximation.

To investigate these effects numerically we will use, like in our previous paper \cite{5}, the CPA technique with the self-energy describing the on-site Coulomb correlation. This 
approximation yields the increase in the size (capacity) and relative broadening for the majority spin band with respect to the minority spin band. This change of the spin band 
shape is additional to the narrowing of majority spin band with respect to the minority spin band, which is coming from the modified inter-site Hartree-Fock approximation. 

The Hartree-Fock approximation is applied to small on-site exchange interaction and small inter-site interactions. As a result the on-site small interactions merely contribute to the 
effective Weiss field, $I_{ex}$ , but the inter-site interactions contribute not only to the Weiss field but also to the spin band narrowing different for both spin directions.

Summarizing, we will have now three driving forces towards the magnetic ordering. 

One is the total exchange interaction, which is the sum of different on-site and inter-site Hartree-Fock interactions. This interaction shifts the majority spin band below the minority 
spin band. Intuitively speaking this interaction is coming from lowering the energy of parallel spins with respect to antiparallel spins [see Eq. (\ref{11}) below] and it follows the 
idea of Slater \cite{7} and Hund \cite{8}. 

The second driving force is the hopping interaction and exchange-hopping interaction, which are lowering the kinetic energy, mentioned above. These interactions in the 
Hartree-Fock approximation also give rise to the shift of spin bands in energy, which is proportional to the magnetization [see Eq. (\ref{10})]. 

The third driving force is the change in shape of the spin band under the influence of strong Coulomb on-site interaction, $U$, and the inter-site interactions \cite{9,10} (in here 
we will use only $\Delta t$ and $t_{ex}$  interactions). These effects of the change in the band shape, helping ferromagnetism, are mathematically described by the so-called 
correlation factors \cite{4,5,11} $K_U$ for $U$ interaction and $K_b$ for $\Delta t$ and $t_{ex}$ interactions.

The hopping interaction and the exchange-hopping interaction, treated in the Hartree-Fock approximation, contribute also to the superconductivity effect within the BCS 
formalism (see e.g. Refs. \cite{12,13,14,15,16}). In particular the exchange-hopping interaction is the driving force towards the d-wave superconductivity \cite{15} and p-wave 
superconductivity \cite{16}. 

The density of state (DOS), previously semi-elliptic \cite{5}, is now assumed to be more realistic; distorted-elliptic, which has the maximum around the Fermi energy (see Wahle et. 
al. \cite{17} with their parameter $a\neq0$  and close to minus one). 

The paper is organized as follows. In Section 2, we have put forward the model Hamiltonian and developed the formalism to treat the on-site and inter-site Coulomb correlation. 
Numerical examples are presented in Section 3 based on the semi-elliptic DOS for the weak and strong Coulomb correlation. In addition we analyze the influence of the 
asymmetrical DOS on the magnitude of the on-site exchange interaction. On the base of these results, the conclusions regarding the appearance of magnetic ordering with 
growing occupation of the band are drawn in Section 4.

\vskip1.0cm 

\noindent {\Large {\bf 2. The Model Hamiltonian}} 

\vskip0.5cm 
 
The Hamiltonian for the one-band Hubbard model can be written in the form\cite{1} 
	\be
 \begin{array}{c}
 H=\left( {T_0 -\mu _0 } \right)\sum\limits_i {\hat {n}_i } 
-\sum\limits_{<ij>\sigma } {t_{ij}^\sigma \left( {c_{i\sigma }^+ c_{j\sigma 
} +h.c.} \right)} -I_{ex} \sum\limits_{i\sigma } {\tilde {n}_\sigma \hat 
{n}_{i\sigma } } +U\sum\limits_{i\sigma } {\hat {n}_{i\sigma } \hat 
{n}_{i-\sigma } } \\ 
 +V\sum\limits_{<ij>} {\hat {n}_i \hat {n}_j } +J\sum\limits_{<ij>\sigma 
,\sigma '} {c_{i\sigma }^+ c_{j\sigma '}^+ c_{i\sigma '} c_{j\sigma } } 
+J'\sum\limits_{<ij>} {\left( {c_{i\uparrow }^+ c_{i\downarrow }^+ 
c_{j\downarrow } c_{j\uparrow } +h.c.} \right)} \\ 
 \end{array},
\label{1}
\ee
where $\mu_0$ is the chemical potential, $c_{i\sigma }^+({c_{i\sigma } })$ creates (destroys) an electron of spin $\sigma$ on the $i$th lattice site, $\hat n_{i\sigma }  = c_{i\sigma 
}^ +  c_{i\sigma }$ is the electron number operator for electrons with spin $\sigma$  on the$i$th lattice site, $\hat n_i  = \hat n_{i\sigma }  + \hat n_{i - \sigma }$ is the operator of 
the total number of electrons on the $i$th lattice site, $\tilde n_\sigma$ is the probability of finding the electron with spin $\sigma$ in a given band, $T_0$ is the local energy level, 
$U$ is the on-site Coulomb repulsion and $I_{ex}$ is the on-site exchange interaction. In the Hamiltonian (1) we have three explicit inter-site interactions; $J$-exchange interaction, 
$J'$-pair hopping interaction, $V$-density-density interaction. The spin dependent correlation hopping $t_{ij}^\sigma$ depends on the occupation of sites $i$ and $j$, and in the 
operator form it can be expressed as  

	\be
t_{ij}^\sigma   = t_0 (1 - \hat n_{i - \sigma } )(1 - \hat n_{j - \sigma } ) + t_1 \left[ {\hat n_{i - \sigma } (1 - \hat n_{j - \sigma } ) + \hat n_{j - \sigma } (1 - \hat n_{i - \sigma } )} 
\right] + t_2 \hat n_{i - \sigma } \hat n_{j - \sigma },
\label{2} 
\ee 

where $t_0$ gives the hopping amplitude for an electron of spin $\sigma$ when both sites $i$ and $j$ are empty. Parameter $t_1$  gives the hopping amplitude for an electron of 
spin $\sigma$  when one of the sites $i$ or $j$ is occupied by an electron with opposite spin. Parameter $t_2$ gives the hopping amplitude for an electron of spin $\sigma$ when 
both sites $i$ and $j$ are occupied by electrons with opposite spin. Quite recently, several authors suggested that the expected relation $t_0>t_1>t_2$, may be reversed for large 
enough inter-atomic distances, $t_0<t_1<t_2$  (see \cite{13} and \cite{18}). This concept would fit to the results of Gunnarsson and Christensen \cite{19}, who for the heavier 
elements (e.g. 3d or 4f) claim growing hopping integrals with increasing occupation.

For the total on-site exchange interaction $I_{ex}$  one can write on the microscopic level the following expression 

	\be
	I_{ex}  = \left( {d - 1} \right)\left( {J_{in}  + J'_{in}  + V_{in} } \right),
	\label{3}
	\ee
where $d$ is the number of sub-bands (orbitals) within the same band, $J_{in}$, $J'_{in}$ and $V_{in}$ are the exchange, pair-hopping, and density-density interactions between 
different orbitals within the same atomic site. In the case of the weak correlation this interaction is augmented by the Coulomb repulsion $U$.

Including the occupationally dependent hopping given by Eq. (\ref{2}) into the Hamiltonian (\ref{1}) we obtain the following result 
	 \be
	\begin{array}{c}
 H=\left( {T_0 -\mu _0 } \right)\sum\limits_i {\hat {n}_i } -I_{ex} 
\sum\limits_{i\sigma } {\tilde {n}_\sigma \hat {n}_{i\sigma } } 
+U\sum\limits_{i\sigma } {\hat {n}_{i\sigma } \hat {n}_{i-\sigma } } 
+V\sum\limits_{<ij>} {\hat {n}_i \hat {n}_j } \\ 
 +J\sum\limits_{<ij>\sigma ,\sigma '} {c_{i\sigma }^+ c_{j\sigma '}^+ 
c_{i\sigma '} c_{j\sigma } } +J'\sum\limits_{<ij>} {\left( {c_{i\uparrow }^+ 
c_{i\downarrow }^+ c_{j\downarrow } c_{j\uparrow } +h.c.} \right)} \\ 
 -\sum\limits_{<ij>\sigma } {\left[ {t_0 -\Delta t\left( {\hat {n}_{i-\sigma 
} +\hat {n}_{j-\sigma } } \right)+2t_{ex} \hat {n}_{i-\sigma } \hat 
{n}_{j-\sigma } } \right]\left( {c_{i\sigma }^+ c_{j\sigma } +h.c.} \right)} 
\\ 
 \end{array},
 \label{4}
\ee	  
where
	\be
	\Delta t = t_0  - t_1   ,\quad  
t_{ex}=\frac{{t_0  + t_2 }}{2} - t_1 .
\label{5}
  \ee  
In this form it is quite visible that $\Delta t$ and $t_{ex}$ are also the inter-site interactions. The Hamiltonian (\ref{4}) will be analyzed in two steps. For the kinetic part (the terms 
with the inter-site interactions $\Delta t$ and $t_{ex}$) the Hartree-Fock approximation will be applied. The other inter-site interactions, $V$, $J$, and $J'$, also treated in the 
Hartree-Fock approximation, will be nonzero in the equations of this section to have a full approach, but later on in the numerical analysis they will be assumed to be zero as to 
limit the number of free parameters. The role of the inter-site interactions, $V$, $J$, and $J'$, was already studied by us before \cite{5}.
For the strong Coulomb repulsion $U$ the CPA will be used. After setting the energy scale at the atomic level $T_0$, and performing the Hartree-Fock approximation we will 
obtain 
	\be
H_{MF}  =  - \sum\limits_{ < ij > \sigma } {t_{eff}^\sigma  \left( {c_{i\sigma }^ +  c_{j\sigma }  + h.c.} \right)}  - \sum\limits_{i\sigma } {(\mu  - M_\sigma  )\hat n_{i\sigma } }  + 
U\sum\limits_{i\sigma } {\hat n_{i\sigma } \hat n_{i - \sigma } } ,
	\label{6}
	\ee
where $t_{eff}^\sigma$ is effective spin-dependent hopping integral given by  
	\be
t_{eff}^\sigma   = t_0 b_\sigma ,
\label{7}  
	\ee 
with the parameter $b_\sigma$ describing the spin dependent change of the bandwidth
	\be
b_\sigma   = 1 - 2\frac{{\Delta t}}{{t_0 }}\tilde n_{ - \sigma }  + 2\frac{{t_{ex} }}{{t_0 }}\left( {\tilde n_{ - \sigma }^2  - I_{ - \sigma }^2  - I_\sigma  I_{ - \sigma } } \right) - 
\frac{{J - V}}{{t_0 }}I_\sigma   - \frac{{J + J'}}{{t_0 }}I_{ - \sigma } .
\label{8} 
\ee 
The parameter $I_\sigma=\left\langle{c_{i\sigma }^+c_{j\sigma }}\right\rangle$ is the average bond occupation for spin $\sigma$ and the quantity $\tilde n_{ - \sigma }$ above is 
the probability of finding the electron with spin $-\sigma$ in a given band. For the weak correlation one can assume that probability $\tilde n_{ - \sigma }$ is equal to the average 
number of electrons with spin $-\sigma$, i.e. $\tilde n_{ - \sigma }=n_{-\sigma}$. In the case of strong correlation $(U\gg D)$  probability of occupation of the band with spin 
$-\sigma$, $\tilde n_{ - \sigma }$, will depend on which from the split Hubbard sub-bands we are in. For the lower sub-band, $\varepsilon\approx 0$ and $n<1$ , we assume that 
$\tilde n_{ - \sigma }=1-n_{-\sigma}$, but for the upper sub-band, $\varepsilon\approx U$ and $n>1$, we have to assume that $\tilde n_{ - \sigma }=n_{-\sigma}$.
The modified spin-dependent chemical potential $\mu$ is given by 
	\be
	\mu =\mu _0 -zVn
	\label{9}
	\ee   	
and  $M_\sigma$ is the spin-dependent modified molecular field expressed as 
	\be
	M_\sigma =-I_{ex} \tilde {n}_\sigma +2z\Delta tI_{-\sigma } -2zt_{ex} 
I_{-\sigma } \tilde {n}_\sigma -zJ\tilde {n}_\sigma ,
\label{10}
	\ee  
with $z$ being the number of nearest-neighbors.	
The expression (\ref{10}) for the total molecular field shows clearly that the difference in hopping integrals $t_{ex}$ plays the same role in creating the exchange field (the third 
term above) as the on-site exchange interaction $I_{ex}$ given by Eq. (\ref{11}). 

In the simple interpretation of Slater \cite{7}, one can understand the total on-site exchange interaction $I_{ex}$  as the interaction lowering the energy of each pair of parallel 
spins with respect to the anti-parallel spins according to the equation
	 \be
	I_{ex} =\frac{I_{++} +I_{--} }{2}-I_{+-} \quad .
	\label{11}
	\ee  
The similarity of the $t_{ex}$ definition [Eq. (\ref{5})] with the intuitive definition of exchange interaction $I_{ex}$ given above, and the very same way they contribute to the 
total exchange field [Eq. (\ref{10})] will allow us to call the quantity $t_{ex}$, the exchange-hopping interaction. As it was mentioned above we may have either $t_0>t_1>t_2$ or 
$t_0<t_1<t_2$, depending on the inter-atomic distance. In some cases the hopping integral may grow with increasing electron occupation, $t_0<t_1<t_2$ (see Ref. \cite{13}), 
since the overlap between nearest-neighbors atomic orbitals grows as they are expanding. We will use both these options below in numerical analyzes of the results. 

For the linear decrease or increase of $t_i$ with the growing occupation $n$, the exchange-hopping interaction from Eq. (\ref{5}) is equal zero. But the dependence $t(n)$ most 
likely is not linear. In the case of ${\rm d}^2 t/{\rm d}n^2 > 0$ we have $t_{ex}<0$, and for ${\rm d}^2 t/{\rm d}n^2 < 0$  we have $t_{ex}>0$. 

After Fourier transform of the kinetic energy in Hamiltonian (\ref{6}) we obtain
	\be
	H_{MF} =\sum\limits_{k\sigma } {E_k^\sigma \hat {n}_{k\sigma } } 
+U\sum\limits_{i\sigma } {\hat {n}_{i\sigma } \hat {n}_{i-\sigma } } ,
\label{12}
	\ee 
where the spin dependent electron dispersion relation is given by 
	\be
	E_k^\sigma =\varepsilon _k b_\sigma -\mu +M_\sigma ,
	\label{13}
	\ee  
with $\varepsilon _k$ being the initial (without interactions) dispersion energy of the electron 
	\be
	\varepsilon _k =-t_0 \gamma _k \quad ,
\quad
\gamma _k =\sum\limits_{<i,j>} {e^{i{\rm {\bf k}}\left( {{\rm {\bf R}}_{\rm 
{\bf i}} -{\rm {\bf R}}_{\rm {\bf j}} } \right)}} .
\label{14}
	\ee 
In the case of strong on-site correlation $U$ the CPA decoupling is used to analyze Hamiltonian (\ref{12}), and is described by the following equation 
 	\be
	(1-n_{-\sigma } )\frac{-\Sigma _\sigma }{1+\Sigma _\sigma F_\sigma 
(\varepsilon )}+n_{-\sigma } \frac{U-\Sigma _\sigma }{1-(U-\Sigma _\sigma 
)F_\sigma (\varepsilon )}=0,
\label{15}
	\ee 
where $\Sigma _\sigma$ is the on-site self-energy and $F_\sigma (\varepsilon)$ is the spin dependent Slater-Koster function given by  
	\be
	F_\sigma (\varepsilon )=\frac{1}{N}\sum\limits_k {\frac{1}{\varepsilon 
-E_k^\sigma -\Sigma _\sigma }} .
\label{16}
	\ee 
This function $F_\sigma (\varepsilon)$ can be expressed by the unperturbed function 
	\be
	F_0 (\varepsilon )=\frac{1}{N}\sum\limits_k {\frac{1}{\varepsilon 
-\varepsilon _k }} ,
\label{17}
	\ee  
with the help of the following relation \cite{5}:
	\be
	F_\sigma (\varepsilon)=\frac{1}{b_\sigma }F_0 \left ({\frac{\varepsilon -M_\sigma +\mu -\Sigma _\sigma }{b_\sigma }}\right ),
\label{18}
	\ee  
which becomes the standard CPA relation when there are no inter-site interactions and in consequence $b_\sigma \equiv 1$.

For the spin dependent electron DOS one may write the usual expression  
	\be
	\rho _\sigma (\varepsilon )=-\frac{1}{\pi }{\rm Im} F_\sigma (\varepsilon ).
	\label{19}
	\ee  
The spin-dependent average occupation number $n_\sigma$ is given by
	\be
	n_\sigma =\int\limits_{-\infty }^{+\infty } {\rho _\sigma (\varepsilon 
)f(\varepsilon )d\varepsilon } ,
\label{20}
	\ee 
where $f(\varepsilon)$ is the Fermi function 
	\be
	f_\sigma (\varepsilon )=\frac{1}{1+\exp [\varepsilon -(\mu -M_\sigma )]},
\quad
\beta =\frac{1}{k_B T}.
\label{21}
	\ee  
For magnetization per atom in Bohr's magnetons we can write that
	\be
	m=n_\sigma -n_{-\sigma } .
	\label{22}
	\ee 
Differentiating Eqs. (\ref{22}) and (\ref{20}), with respect to $m$ and assuming later that $m\rightarrow 0$ one obtains the criterion for the ferromagnetic state in the following form
	\be
	1=K+2\left( {\left. {\frac{\partial M_\sigma }{\partial m}} \right|_{m\to 0} 
} \right)\int\limits_{-\infty }^{+\infty } {\rho (\varepsilon 
)f^2(\varepsilon )\exp \left[ {\beta \left( {\varepsilon -\mu } \right)} 
\right]d\varepsilon } ,
\label{23}
	\ee  
where $\rho(\varepsilon)$ is the paramagnetic limit of $\rho _\sigma(\varepsilon)$, the correlation factor $K$ describes the role of change in band shape for creating magnetization 
and it is the sum of the on-site and inter-site correlation factors
	\be
	K=K_U +K_b ,
	\label{24}
	\ee  
where
	\be
	K_U =-\frac{2}{\pi } {\rm Im}\int\limits_{-\infty }^{+\infty } {\frac{\partial 
F_\sigma (\varepsilon )}{\partial \Sigma _\sigma }\frac{\partial \Sigma 
_\sigma }{\partial m}} f(\varepsilon )d\varepsilon
\label{25} 
	\ee 
and
	\be
	K_b =-\frac{2}{\pi }{\rm Im}\int\limits_{-\infty }^{+\infty } {\frac{\partial 
F_\sigma (\varepsilon )}{\partial b_\sigma }\frac{\partial b_\sigma 
}{\partial m}} f(\varepsilon )d\varepsilon .
\label{26}
	\ee  
At zero temperature we obtain from Eq. (\ref{23}) using Eqs. (\ref{24})-(\ref{26}), and (\ref{10}) the following dependence of the critical on-site exchange interaction on carrier 
concentration 
	\be
	I_{ex}^{cr} =\frac{1-K_U -K_b }{\rho (\varepsilon _F )}-\left( {zJ+I_{\Delta 
t} +I_{tex} } \right),
\label{27}
	\ee  
	     
	\be
	I_{\Delta t} =-4z\Delta t\left. {\frac{\partial I_{-\sigma } }{\partial m}} 
\right|_{m\to 0}\qquad \textrm{and} \qquad I_{t_{ex}} =2zt_{ex} \left( {I_0 +n\left. 
{\frac{\partial I_{-\sigma } }{\partial m}} \right|_{m\to 0} } \right),
\label{28}
\ee 
	
where $I_0$ is the average band occupation in the paramagnetic state. The last term on the right hand side of equation (\ref{27}) is the sum of exchange fields coming from 
inter-site exchange interaction; $zJ$, hopping interaction, $I_{\Delta t}$, and the exchange-hopping interaction; $I_{t_{ex}}$.
The parameter of the bandwidth change $b_\sigma$ and the modified molecular field $M_\sigma$ depend on the quantity $I_\sigma$, the average bond occupation for spin 
$\sigma$, which can be expressed as 
	\be
	I_\sigma =<c_{i\sigma }^+ c_{j\sigma } >=\sum\limits_k {\exp [i{\rm {\bf 
k}}({\rm {\bf R}}_{\rm {\bf i}} -{\rm {\bf R}}_{\rm {\bf j}} )]} 
\int\limits_{-\infty }^{+\infty } {f(\varepsilon )S_{k\sigma } \left( 
\varepsilon \right)d\varepsilon } ,
\label{29}
	\ee  
where $S_{k\sigma}(\varepsilon)$ is the single-electron spectral density
	\be
	\begin{array}{c}
 S_{k\sigma } (\varepsilon )=-\frac{1}{\pi }{\rm Im} <<c_{k\sigma }^+ ;c_{k\sigma } >>
 =-\frac{1}{\pi }\frac{{\rm Im} \Sigma _\sigma (\varepsilon )}{\left[ {\varepsilon 
-E_k^\sigma -{\rm Re}\Sigma _\sigma (\varepsilon )} \right]^2+\left[ {{\rm Im} \Sigma 
_\sigma (\varepsilon )} \right]^2} \\ 
 \end{array}.
 \label{30}
	\ee 
On the other side we can get from the kinetic part of the Hamiltonian with spin $\sigma$, $K^\sigma$, that its average value is
	\be
	\left\langle {K^\sigma } \right\rangle =-t_{eff}^\sigma \sum\limits_{<i,j>} 
{\left\langle {c_{i\sigma }^+ c_{j\sigma } } \right\rangle } 
=-zt_{eff}^\sigma I_\sigma ,
\label{31}
	\ee  
or directly from the definition of the average kinetic energy
	\be
	\left\langle {K^\sigma } \right\rangle =\int\limits_{-D_0 }^{\mu _0^\sigma } 
{f({E_\sigma })E_\sigma \rho _\sigma ({\varepsilon _0 } 
)d\varepsilon _0 } ,
\quad
D_0 =zt_0 ,
\label{32}	\ee    
where 
	\be
	E_\sigma (\varepsilon _0 )=\varepsilon _0 b_\sigma ,
\quad
\varepsilon _0 \equiv \varepsilon _k .
\label{33}
	\ee 
Comparing expression (\ref{31}) and (\ref{32}) we have
	\be
	I_\sigma =-\frac{1}{D_\sigma }\int\limits_{-D_0 }^{\mu _0^\sigma } {f\left( 
{E_\sigma } \right)E_\sigma \rho _\sigma \left( {\varepsilon _0 } 
\right)d\varepsilon _0 } ,
\quad
D_\sigma =D_0 b^\sigma .
\label{34}
	\ee  	
We can compare now Eqs. (\ref{29}) and (\ref{34}). They are equivalent when ${\mathop{\rm Re}\nolimits} \Sigma _\sigma  \left( \varepsilon  \right) \Rightarrow 0$ and 
${\mathop{\rm Im}\nolimits} \Sigma _\sigma  \left( \varepsilon  \right) \Rightarrow 0^+$ (we remind here that $E_\sigma  /D_\sigma   = \varepsilon _0 /D_0 \equiv \gamma _k$). 
This means that $I_\sigma$  from Eq. (\ref{29}) is the generalization of the average bond occupation for spin $\sigma$ from Eq. (\ref{34}) to the case of interaction being described 
in the single site approximation by the self-energy $\Sigma (\varepsilon)$.

\vskip1.0cm 

\noindent {\Large {\bf 3. Numerical results and discussion}} 

\vskip0.5cm

The features of this new model will be illustrated by showing the dependence of the critical on-site exchange interaction on the carrier concentration $I_{ex}^{cr}(n)$. 

For the unperturbed DOS we will use the asymmetrical function \cite{17}

\be
\rho _0 \left( \varepsilon \right)=c\frac{\sqrt {D_0 ^2-\varepsilon ^2} 
}{D_0 +a\varepsilon },
\quad
c=\frac{1+\sqrt {1-a^2} }{\pi D_0 },
\label{35}
\ee  
	 
where $D_0$ is the unperturbed half-bandwidth and $a$ is the asymmetry parameter. At $a=0$ this will become the semi-elliptic DOS, and for $a=1$ we obtain the DOS for the 3-d 
fcc lattice with $t'=t_0/4$, where $t_0$- nearest-neighbor hopping and $t'$ the next-nearest-neighbor hopping \cite{20}.

The Slater-Koster function corresponding to this DOS can be calculated from the following formula \cite{3}

\be
F_0 (\varepsilon )=\int\limits_{-\infty }^{+\infty } {\rho _0 (\varepsilon 
')\frac{d\varepsilon '}{\varepsilon -\varepsilon '}} ,
\label{36}
\ee

what will result in the equation

\be
F_0 (\varepsilon )=\frac{c\pi D_0 }{D_0 +a\varepsilon }\left[ 
{\frac{\varepsilon }{D_0 }-\sqrt {\left( {\frac{\varepsilon }{D_0 }} 
\right)^2-1} +\frac{1}{a}-\sqrt {\left( {\frac{1}{a}} \right)^2-1} } 
\right].
\label{37}
\ee

At $a=0$ this will give the well-known result for the semi-elliptic DOS

\be
F_0(\varepsilon)=\frac{2}{D_0 }\left[ {\frac{\varepsilon 
}{D_0 }-\sqrt {\left( {\frac{\varepsilon }{D_0 }} \right)^2-1} } \right],
\label{38}
\ee  
	
which will be used initially. 

In analyzing our model we will concentrate mainly on how the hopping interaction $\Delta t$ and the exchange-hopping interaction $t_{ex}$ influence the ferromagnetic state. We 
will assume all other inter-site interactions; $J=J'=V\equiv 0$. Therefore, the bandwidth factor $b_\sigma$ [from Eq. (\ref{8})], the chemical potential and the spin-dependent 
modified molecular field will simplify to

\be
b_\sigma =1-2\frac{\Delta t}{t_0 }\tilde {n}_{-\sigma } +2\frac{t_{ex} }{t_0 
}\left( {\tilde {n}_{-\sigma }^2 -I_{-\sigma }^2 -I_\sigma I_{-\sigma } } 
\right),
\label{39}
\ee

\be
\mu=\mu _0-T_0,
\label{40}
\ee

\be
M_\sigma =-I_{ex} \tilde {n}_\sigma +2z\Delta tI_{-\sigma } -2zt_{ex} 
I_{-\sigma } \tilde {n}_\sigma .
\label{41}
\ee

The sign and magnitude of interactions $\Delta t$ and $t_{ex}$, according to Eq. (\ref{5}), depend on the hopping amplitudes $t_{0}$, $t_1$, and $t_2$. We assume that 
$t_1/t_0=S$ and $t_1/t_2=S_1$. In general these parameters are different and they both fulfill the condition $S<1$ and $S_1<1$ what is equivalent to $t_{0}>t_1>t_2$ (see Ref. 
\ref{13}). However, in his paper Hirsch \cite{13} has pointed out that for the hydrogen molecule H$_2$ these integrals depend strongly on the inter-atomic distance and for the 
distance large enough we can even have the reverse relation $t_{0}<t_1<t_2$ . The heavier elements (e.g. 3d or 4f) posses larger inter-atomic distances, therefore they may have 
growing hopping integrals with increasing occupation. Gunnarsson and Christensen \cite{19} observed such the dependence for 4f transition elements. In analyzing the influence 
of interactions $\Delta t$ and $t_{ex}$ on magnetism we will consider both negative and positive values. Taking into account the above defined relations between hopping 
integrals we can write that $\Delta t = t_0 (1 - S)$ and $t_{ex}  = t_0 \left( 1 + SS_1  - 2S \right)/2$. 

According to the Eqs. (\ref{8}) and (\ref{10}) interactions $\Delta t$ and $t_{ex}$ create ferromagnetism by changing the relative widths of the spin bands, and by shifting them 
with respect to each other. The Stoner-Wohlfarth criterion tells us that the large DOS on the Fermi level also helps ferromagnetism. In the model with symmetrical semi-elliptic DOS 
the large value of the DOS on the Fermi level can be achieved by decreasing the bandwidth. Another important effect associated with changing spin bandwidths is the change of 
the ratio $b_\sigma /b_{-\sigma}$. The ratio $b_\sigma /b_{-\sigma}>1$ helps ferromagnetism for concentrations smaller then half-filling, but for concentrations larger then 
half-filling it is the ratio $b_\sigma /b_{-\sigma}<1$, which helps ferromagnetism \cite{5,9}. The third factor helping ferromagnetism is shifting by interactions the band $-\sigma$ 
to higher energies then the band $+\sigma$.

\begin{figure}[b]
\epsfig{file=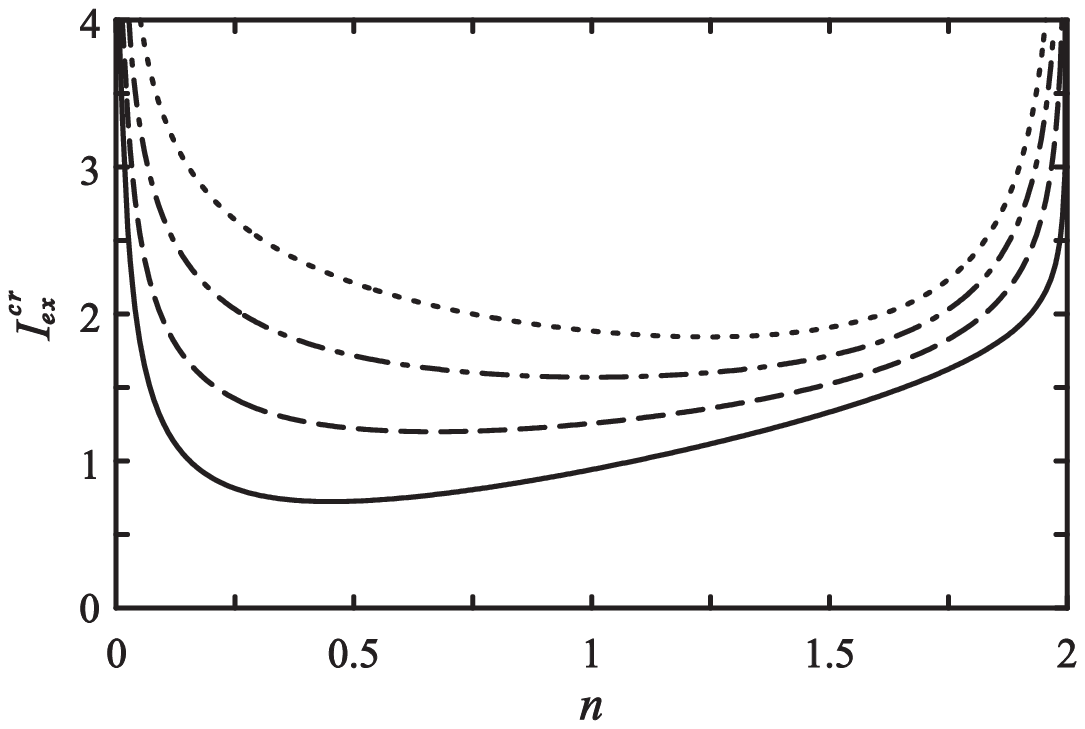, width=0.45\hsize}
 \hspace{0.05\hsize}
\epsfig{file=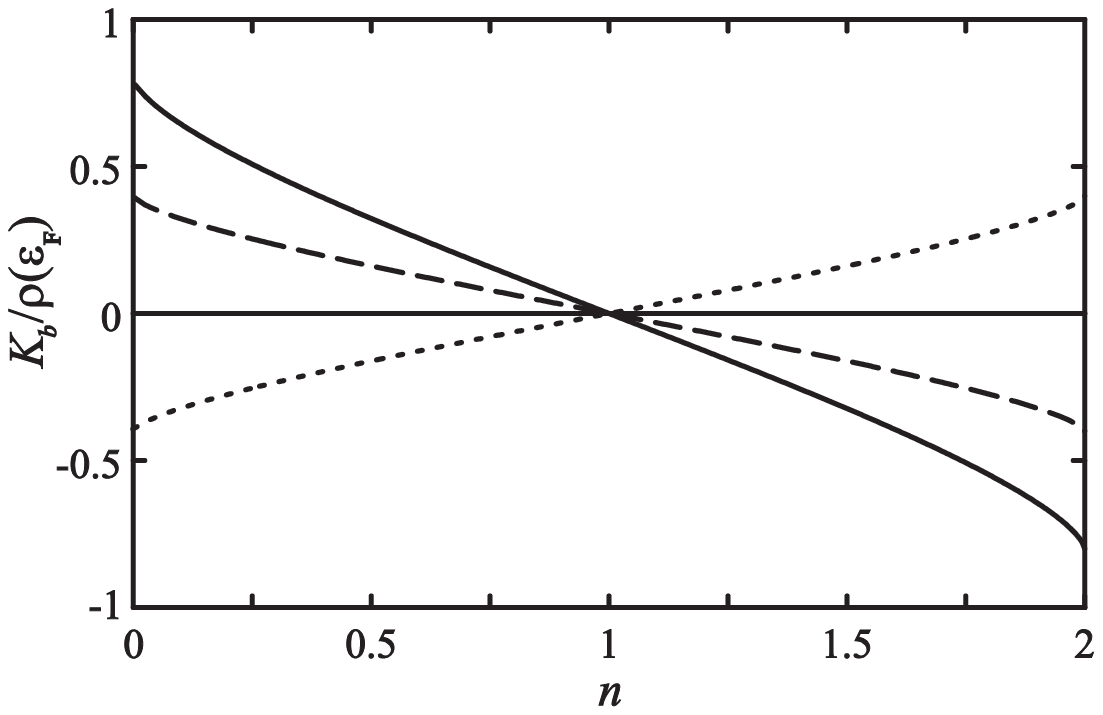,width=0.45\hsize}

 \par\vspace{1.5ex}\makebox[0.5\hsize]
    {\small  FIG. 1(a)}\makebox[0.5\hsize]{\small  FIG. 1(b)}

\vspace{1.5ex}
\epsfig{file=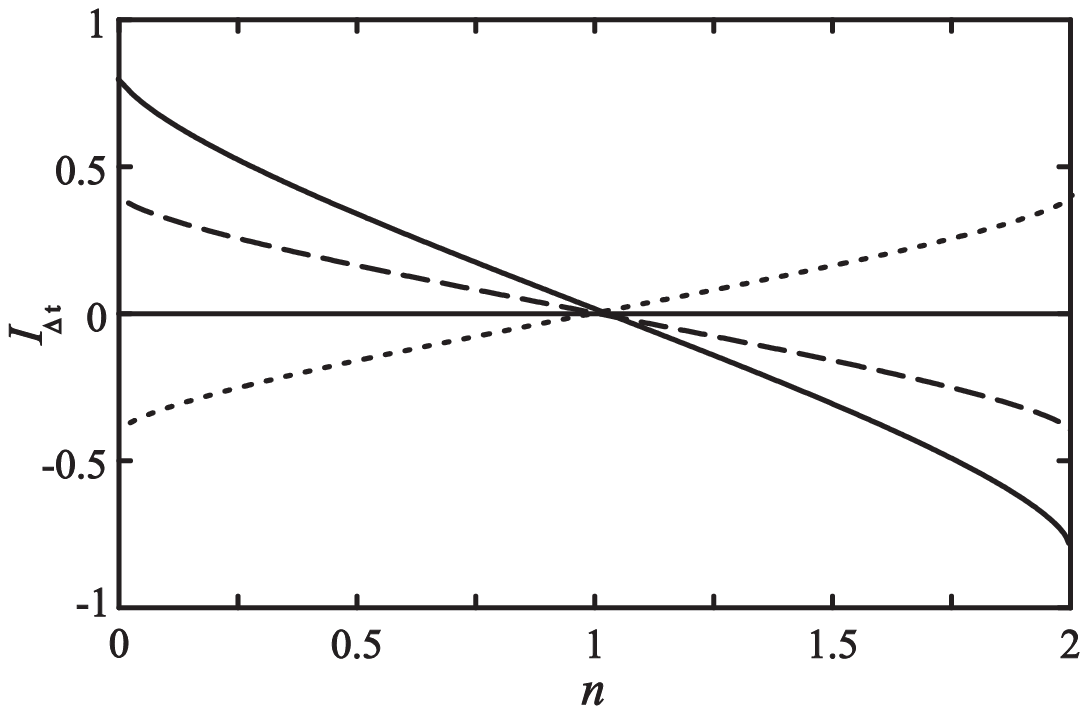, width=0.45\hsize}
 \hspace{0.05\hsize}
\epsfig{file=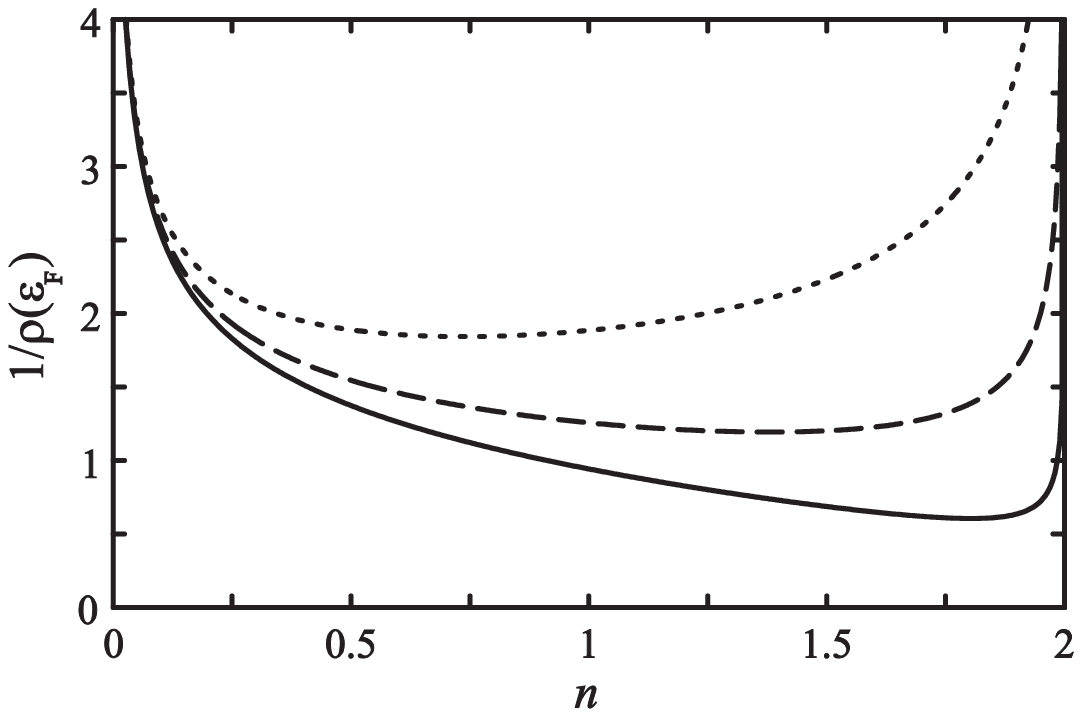, width=0.45\hsize}
\par\vspace{1.5ex}\makebox[0.5\hsize]
    {\small  FIG. 1(c)}\makebox[0.5\hsize]{\small  FIG. 1(d)}
	\vspace{1.5ex}
\caption{ Critical on-site exchange interaction   (FIG. 1(a)),
inter-site correlation factor $K_b/{\rho (\varepsilon _F)}$  (FIG. 1(b)),
  band shift factor $I_{\Delta t}$  (FIG. 1(c)) and DOS reciprocity $1/{\rho (\varepsilon _F)}$ (FIG. 1(d)) versus carrier concentration, at $t_{ex}=0$ and different values of hopping 
interaction; $\Delta t=0.4 t_0$ - solid curve,    $\Delta t=0.2 t_0$ - dashed curve,  $\Delta t=-0.2 t_0$ - dotted curve and $\Delta t=0 t_0$ - dot-dashed curve. All curves in units of 
$D_0$ }
\end{figure}

In Figs. 1 we illustrate the role of hopping interaction; $\Delta t\neq 0$ and $t_{ex}=0$, but in Figs. 2 we have the reverse situation;  $\Delta t=0$ and $t_{ex}\neq 0$. We will start 
the analysis with the weak correlation ($U\ll D)$. For the weak correlation one can assume that probability $\tilde n_{ - \sigma }$ is equal to the average number of spin $-\sigma$ 
electrons, i.e. $\tilde n_{ - \sigma }=n_{ - \sigma }$. For the weak correlation the on-site correlation factor $K_U$ is equal zero. Fig. 1(a) show the dependence $I_{ex}^{cr}(n)$  for 
different values of hopping interaction $\Delta t$ and $t_{ex}=0$.  Analyzing this dependence one can see that the negative value of hopping interaction ($S>1$) increases the 
critical on-site exchange interaction $I_{ex}^{cr}$ above the Stoner-Wohlfarth level (the dot-dashed curve). Positive hopping interaction $\Delta t$ ($S<1$) depletes significantly 
this field especially for electron concentrations below the half-filling. According to Eq. (\ref{27}) hopping interaction modifies the critical on-site exchange interaction through the 
correlation factor $K_b$, the factor $I_{\Delta t}$, and the change in the DOS $\rho (\varepsilon)$. These factors are shown in function of electron concentration for different 
parameters  $\Delta t=0$ and at $t_{ex}\neq 0$ in Figs. 1(b)-1(d). The correlation factor $K_b$ is related to the ratio $b_\sigma /b_{-\sigma}$. According to the Eq. (\ref{27}) to 
enhance the ferromagnetism we need positive $K_b$, which for $n<1$ we obtain at $b_\sigma /b_{-\sigma}>1$, and for $n>1$ at $b_\sigma /b_{-\sigma}<1$. Analyzing $K_b(n)$, 
shown in Fig. 1(b), we see that $K_b$ is positive for small concentrations and is negative for $n>1$. This gives from Eq. (\ref{8}) positive $b_\sigma /b_{-\sigma}>1$ at all 
concentrations. Factor $I_\Delta t$ shown in Fig. 1(c) gives the spin band shift. For small concentrations and $\Delta t>0$ this factor is positive and the band $+\sigma$ is shifted 
lower in energy then the band $-\sigma$ what helps the ferromagnetism. At $n>1$ the factor $I_\Delta t$ is negative, in the result the hopping interaction will oppose there the 
ferromagnetic ordering. The dependence of $1/{\rho (\varepsilon)}$  on concentration shown in Fig. 1(d) has the minimum, which shifts towards larger concentrations with 
growing hopping interaction. Decreasing the bandwidth with growing $n$ (see Eq. (\ref{8})) what increases the DOS causes this effect. 

\begin{figure}[t]
\epsfig{file=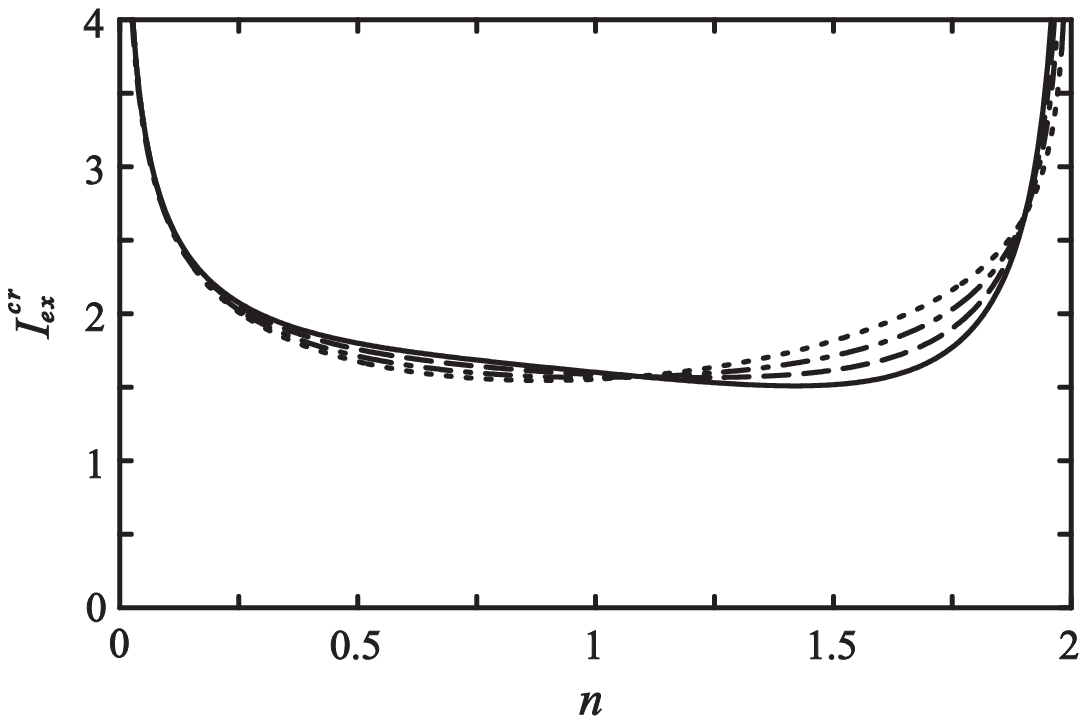, width=0.45\hsize}
 \hspace{0.05\hsize}
\epsfig{file=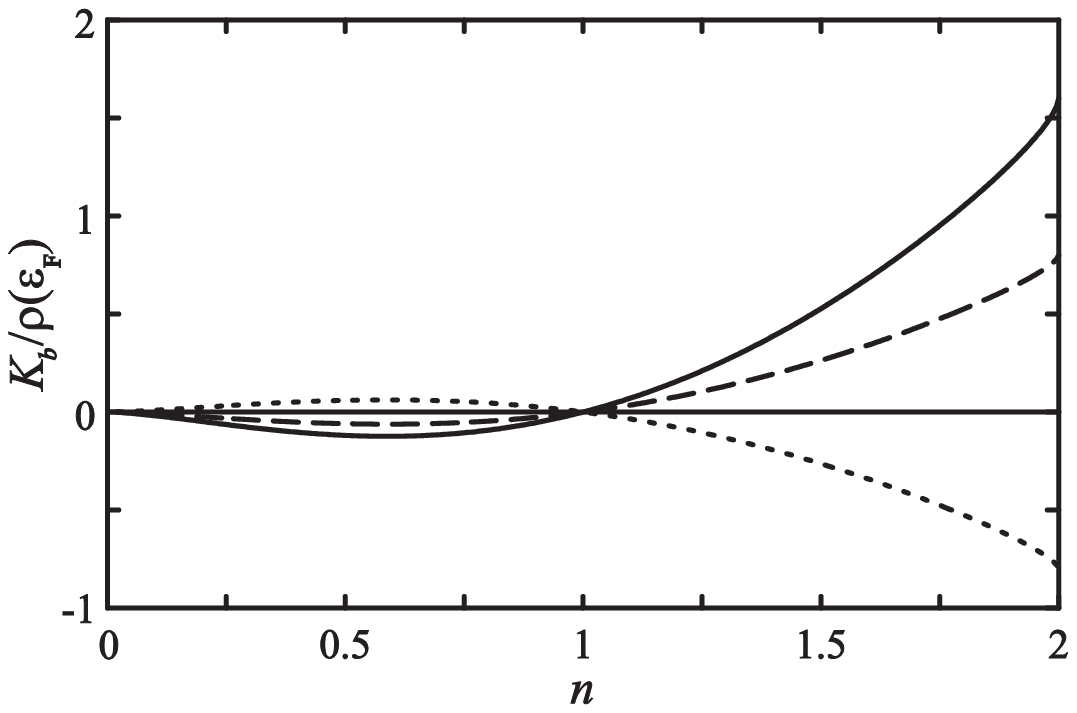,width=0.45\hsize}

 \par\vspace{1.5ex}\makebox[0.5\hsize]
    {\small FIG. 2(a)}\makebox[0.5\hsize]{\small  FIG. 2(b)}

\vspace{1.5ex}
\epsfig{file=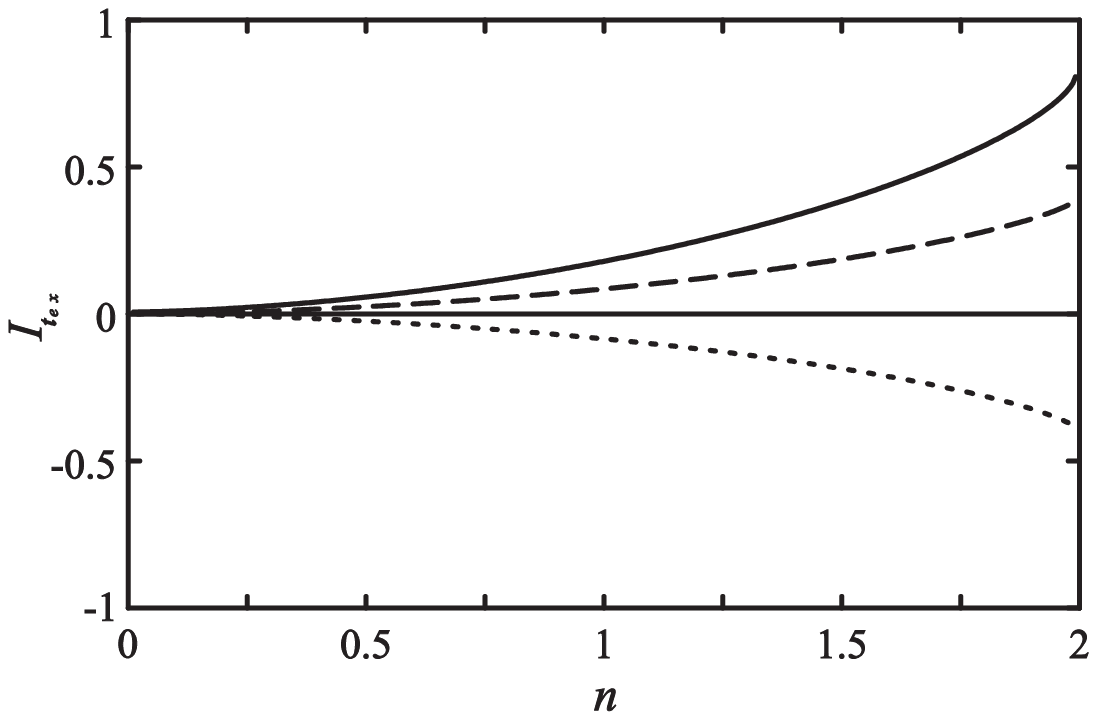, width=0.45\hsize}
 \hspace{0.05\hsize}
\epsfig{file=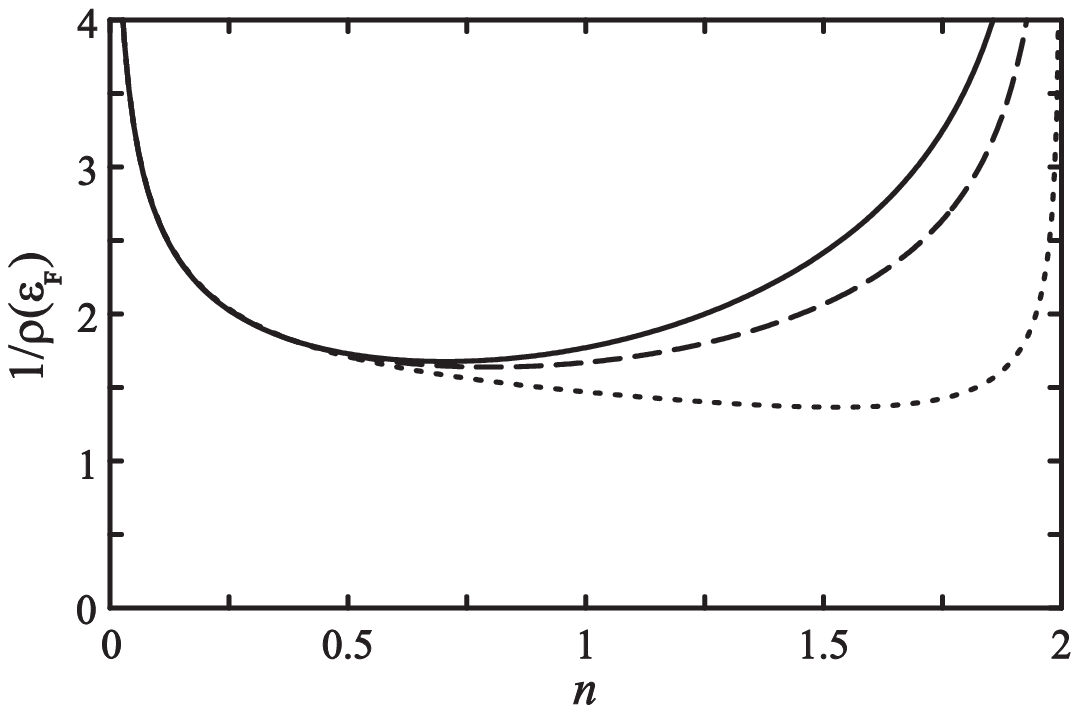, width=0.45\hsize}
\par\vspace{1.5ex}\makebox[0.5\hsize]
    {\small  FIG. 2(c)}\makebox[0.5\hsize]{\small  FIG. 2(d)}
\vspace{0.5ex}
\caption{Critical on-site exchange interaction   ( FIG. 2(a)), inter-site correlation factor $K_b/{\rho (\varepsilon _F)}$  ( FIG. 2(b)),
  band shift factor $I_{t_{ex}}$  ( FIG. 2(c)) and DOS reciprocity $1/{\rho (\varepsilon _F)}$ ( FIG. 2(d)) versus carrier concentration, at $\Delta t=0$ and different values of 
exchange-hopping interaction; $t_{ex}=0.4 t_0$ - solid curve,    $t_{ex}=0.2 t_0$ - dashed curve, $t_{ex}=-0.2 t_0$ - dotted curve and $t_{ex}=0$ - dot-dashed curve. All curves in 
units of $D_0$}
	\end{figure}

Fig. 2(a) presents $I_{ex}^{cr}(n)$ for different values of the exchange-hopping interaction  $t_{ex}$ and $\Delta t=0$. Analyzing those curves one can see that the 
exchange-hopping interaction $t_{ex}$ does not decrease much the critical on-site exchange interaction but rather shifts the minimums (for positive $t_{ex}$) towards larger 
concentrations ($n>1$). Interaction $t_{ex}$ modifies value of $I_{ex}^{cr}$ through the inter-site correlation factor $K_b$, which is related with the spin dependent band shift 
factor $I_{t_{ex}}$ and the DOS reciprocity $1/{\rho (\varepsilon)}$ (see Eq. (\ref{27})). Their dependence on concentration is shown in Figs. 2(b)-2(d). Those curves show that 
for small concentrations the influence of $t_{ex}$ on ferromagnetism is relatively weak. At concentrations larger then half-filling and for positive $t_{ex}$ we obtain the positive 
inter-site correlation factor $K_b$ and positive factor $I_{t_{ex}}$ which enhances ferromagnetism. This enhancement is reduced by the decrease of DOS on the Fermi, which at 
positive interaction $t_{ex}$ comes from the increase in bandwidth with concentration (see Eq. (\ref{8})). 
To recapitulate the hopping interaction $\Delta t$ enables the ferromagnetism for electron concentrations $n<1$, while for $n>1$ its influence is smaller. The exchange-hopping 
interaction $t_{ex}$ helps ferromagnetism for large concentrations. For $n<1$ its effect on ferromagnetism gets weaker.

\begin{figure}[t]
\epsfig{file=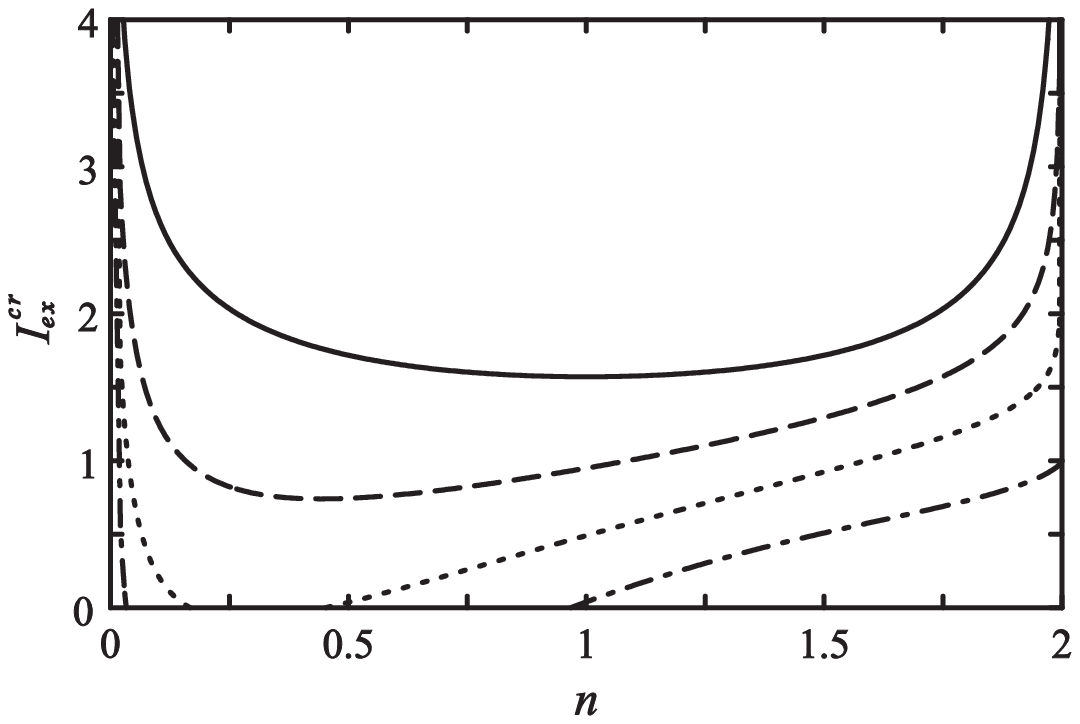, width=0.45\hsize}
 \hspace{0.05\hsize}
\epsfig{file=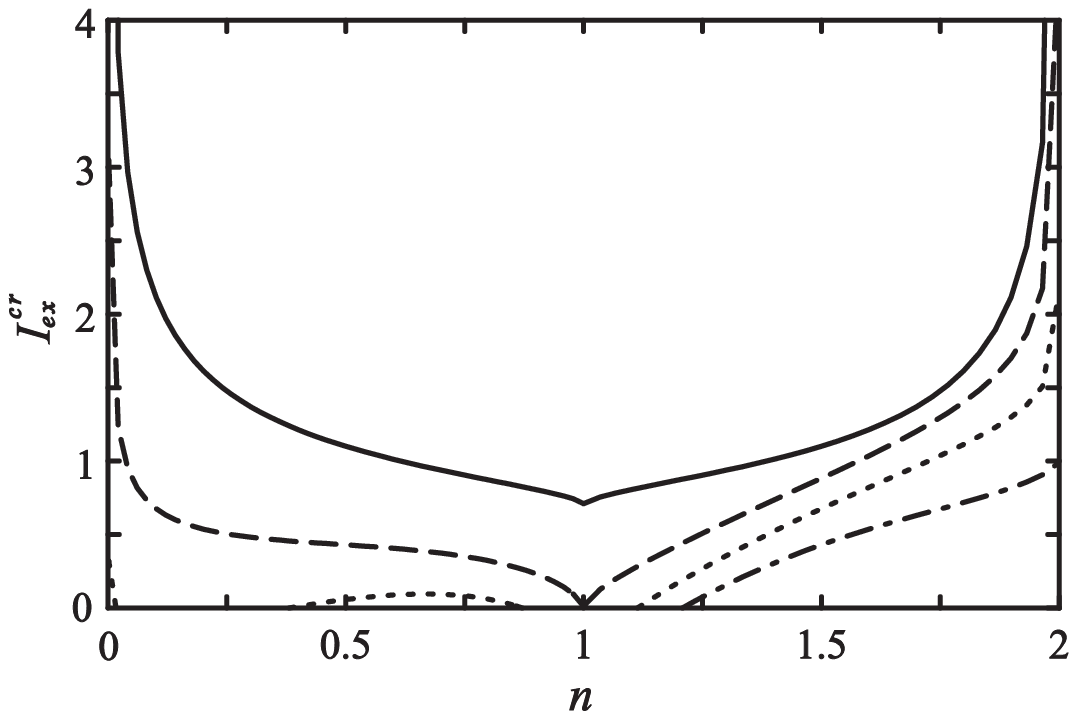,width=0.45\hsize}

 \par\vspace{1.5ex}\makebox[0.5\hsize]
    {\small FIG. 3}\makebox[0.5\hsize]{\small FIG. 4}
	\vspace{0.5ex}
	\caption{ Dependence of the critical on-site exchange interaction (in units of $D_0$) on carrier concentration for the weak Coulomb correlation $U$ and different values of the 
parameter $S$ ($S_1=S$); $S=1$ - solid curve, $S=0.6$ - dashed curve, $S=0.3$ - dotted curve and $S=0$ - dot-dashed curve.}
\vspace{1.5ex}
\caption{ Dependence of the critical on-site exchange interaction (in units of $D_0$) on carrier concentration for the strong Coulomb correlation $U$ and different values of the 
parameter $S$ ($S_1=S$); The curves description is as in FIG. 3}
	\end{figure}
	
 In Figs. 3 and 4 we show the dependence of the critical on-site exchange interaction versus the carrier concentration $I_{ex}^{cr}(n)$ for the weak and strong Coulomb 
correlation $U$. For the curves of $I_{ex}^{cr}(n)$ shown in Figs. 3 and 4 we assumed that $S=S_1$ \cite{18}. In effect we obtained $\Delta t = t_0 (1 - S)$ and $t_{ex}={{t_0(1 - 
S)^2}\mathord{\left/ {\vphantom {{t_0 (1 - S)^2 } 2}} \right. \kern-\nulldelimiterspace} 2}$. 

The curves presented in Figs. 3 and 4 show that the hopping interaction $\Delta t$ together with the exchange-hopping interaction $t_{ex}$ decreases the minimum on-site 
exchange interaction $I_{ex}^{cr}$ necessary for magnetic ordering. At small enough values of the parameter $S$ we obtain the ferromagnetic state for some carrier concentration 
already at the zero values of the on-site exchange interaction $I_{ex}$. The inter-site correlation factor $K_b$ depends on parameter of the bandwidth change $b_\sigma$, which 
in turn is a function of the occupation probability of the band with opposite spin, $\tilde n_{-\sigma}$. As it was already mentioned, in the case of the weak correlation $(U\ll D)$  
one can assume that this probability, $\tilde n_{-\sigma}$, is equal to the average number of spin $-\sigma$ electrons, i.e. $\tilde n_{-\sigma}=n_{-\sigma}$. In the case of strong 
correlation $(U\gg D)$ the probability of occupation of the band with spin $-\sigma$, $\tilde n_{-\sigma}$, will depend on which from the split Hubbard bands we are in. For the 
lower sub-band $(\varepsilon \approx 0)$, when concentration $n<1$ , we assume that  $\tilde n_{-\sigma}=1-n_{-\sigma}$, but for the upper sub-band $(\varepsilon \approx U)$ 
and $n>1$ we have to assume that $\tilde n_{-\sigma}=n_{-\sigma}$. For both weak and strong correlation we obtain minimum of $I_{ex}^{cr}(n)$ for concentrations $n\leq 1$. 
This shows the large influence on ferromagnetism of hopping interaction for the hopping integrals fulfilling the relation $t_0>t_1>t_2$. Therefore, the elements showing 
ferromagnetism at higher concentrations $(n>1)$ should have large difference between $S$ and $S_1$. 

The large DOS, which favors ferromagnetism according to the Stoner-Wohlfarth criterion, can be achieved by decreasing the bandwidth as it was already discussed, but also by 
considering the asymmetrical DOS. To illustrate the results we will use later on the asymmetrical function of  Ref. \cite{17} for the unperturbed DOS [see Eq. (\ref{35})].

\begin{figure}[t]
\begin{center}
\epsfig{file=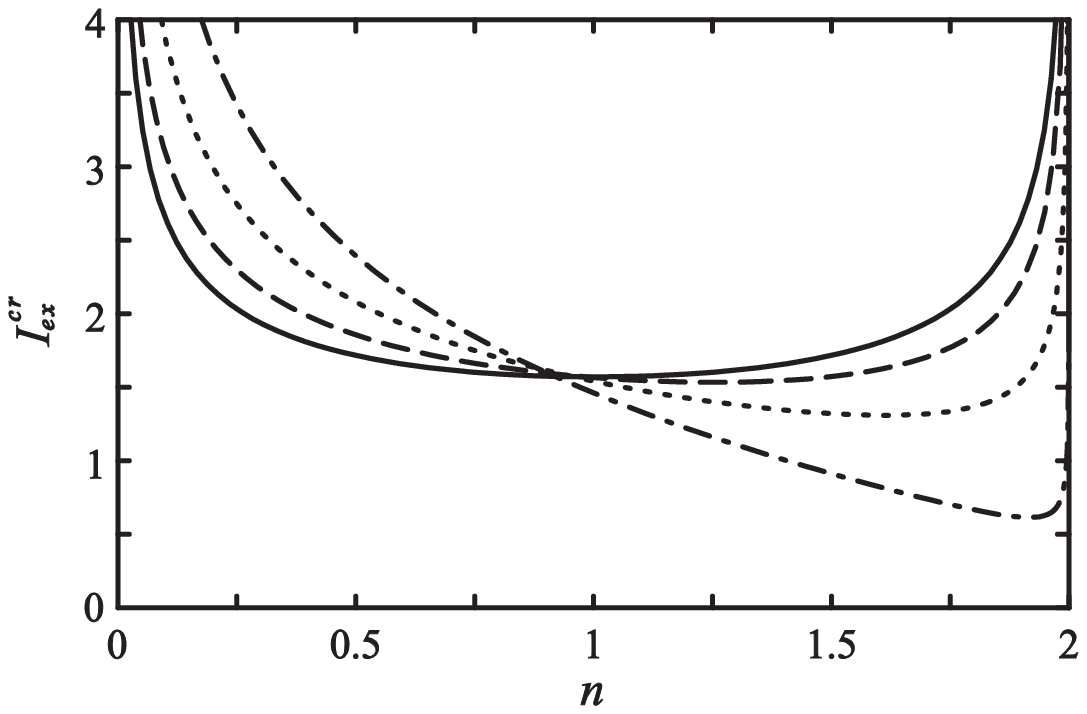, width=0.47\hsize}
\end{center}
\caption{Dependence of the critical on-site exchange interaction (in units of $D_0$) on carrier concentration for the weak Coulomb correlation $a$ and different values of the 
asymmetry parameter; $a=0$ - solid curve, $a=-0.3$ - dashed curve, $a=-0.7$ - dotted curve and $a=-0.97$ - dot-dashed curve.}
\end{figure}

Fig. 5 presents the dependence $I_{ex}^{cr}(n)$ for different values of the asymmetry parameter $a$ in the case of the weak Coulomb correlation $U$. For these curves we 
assumed that $\Delta t=t_{ex}=0$ which corresponds to the hopping integral being independent from the occupation of the sites involved in hopping; $t_0=t_1=t_2$. Analyzing 
these results and comparing them to the corresponding results obtained for symmetrical DOS one can see that the new DOS causes strong asymmetry of the calculated function 
$I_{ex}^{cr}(n)$ $(I_{ex}^{cr}(n)\neq I_{ex}^{cr}(2-n))$. For the negative values of the parameter $a$, the minimum of critical on-site exchange interaction $I_{ex}^{cr}$ will 
appear at concentrations above the half-filling. At positive $a$ this minimum will be shifted to the concentrations smaller then the half-filling. 

The use of the asymmetry parameter increases the DOS at its maximum, what allows for lowering the values of the critical on-site exchange interaction $I_{ex}^{cr}$ as compared 
to the results for the symmetrical semi-elliptic DOS.

\begin{figure}[bp]
\epsfig{file=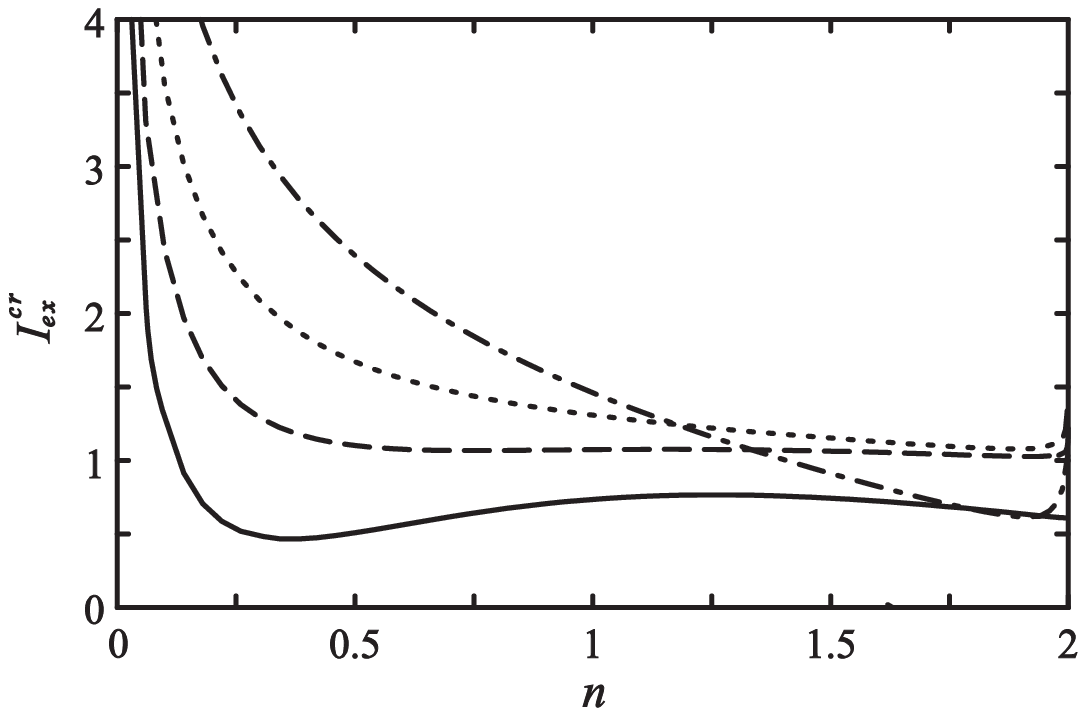, width=0.45\hsize}
 \hspace{0.05\hsize}
\epsfig{file=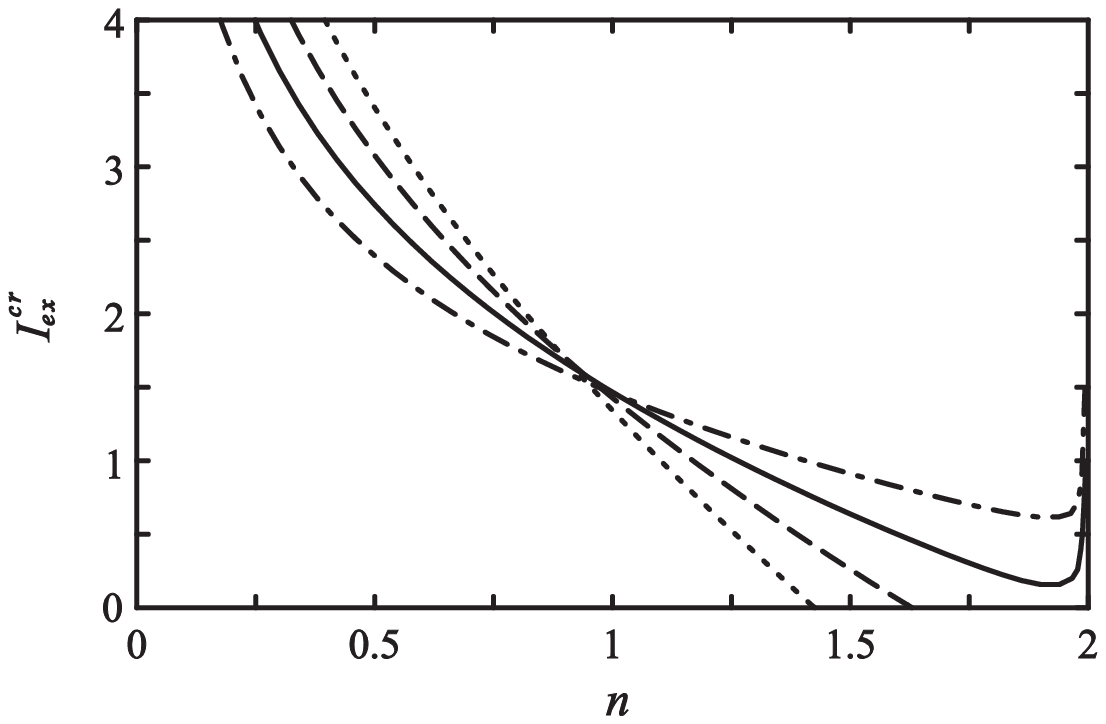,width=0.45\hsize}

 \par\vspace{1.5ex}\makebox[0.5\hsize]
    {\small FIG. 6}\makebox[0.5\hsize]{\small FIG. 7}
	\vspace{1.5ex}
\caption{ Dependence of the critical on-site exchange interaction (in units of $D_0$) on carrier concentration for the weak Coulomb correlation $U$ and the asymmetry parameter 
$a=-0.97$. The values of the parameter   $S$ ($S_1=S$) are below one corresponding to $t_0>t_1>t_2$; $S=0$ - solid curve, $S=0.3$- dashed curve, $S=0.6$ - dotted curve and 
$S=1$ - dot-dashed curve.}
\vspace{1.5ex} 
\caption{ Dependence of the critical on-site exchange interaction (in units of $D_0$) on carrier concentration for the weak Coulomb correlation  $U$ and the asymmetry parameter 
$a=-0.97$. The values of the parameter   $S$ ($S_1=S$) are above one corresponding to $t_0<t_1<t_2$; $S=1.2$ - solid curve, $S=1.4$ - dashed curve, $S=1.6$ - dotted curve and 
$S=1$ - dot-dashed curve.}
\end{figure}

Fig. 6 shows the dependence of $I_{ex}^{cr}(n)$ for $a=-0.97$ and for different values of the parameter $S$ ($S\leq 1$ and we assume that $S_1=S$). One can see that for small 
concentrations ($n<1$) the decrease of $S$ causes decrease in values of the critical on-site exchange interaction $I_{ex}^{cr}$. Small $S$ corresponds to large hopping and 
exchange-hopping interactions, $\Delta t$ and $t_{ex}$, which in this case help the ferromagnetism. In a result for small concentrations we have similar behavior as in the case of 
the semi-elliptic DOS (see Fig. 3). The difference is that the small DOS on the Fermi level (for $a=-0.97$) requires for ferromagnetism the nonzero value of the on-site exchange 
interaction $I_{ex}$. 

For large concentrations ($n>1$ ) and small $S$ ($S<0.5$) the increase of $S$ causes the increase of the critical on-site exchange interaction $I_{ex}^{cr}$ or weakening of 
ferromagnetism. At larger values of $S$ $(0.5<S\leq 1)$the increase of $S$ causes decrease of $I_{ex}^{cr}$. When the parameter $S$ is high enough ($S>1$, see Fig. 7) we can get 
at some concentrations the ferromagnetic state without the on-site exchange interaction; $I_{ex}^{cr}=0$. Such a value of $S$ corresponds to the negative hopping interaction 
$\Delta t$ and positive exchange-hopping interaction $t_{ex}$. For these values of interactions, at $n>1$, both the band shift and the inter-site correlation factor $K_b$ are in 
favor of ferromagnetism. In the case of semi-elliptic DOS the ferromagnetic state is not created at these $n$, $\Delta t$, and $t_{ex}$, since the DOS is too small on the Fermi level. 
The asymmetric DOS (at $a=-0.97$) has large values of $\rho (\varepsilon _F )$, when band is close to full filling, this is why $I_{ex}^{cr}$ drops even to zero. Parameter $S>1$ 
corresponds to $t_0<t_1<t_2$. Such a relation according to Hirsch \cite{13}, can take place for elements with large inter-atomic distances, perhaps elements of 3d and 4f groups. 
These elements show also strong asymmetry in DOS with high density on the Fermi level at the end of the 3d or 4f row. The model presented in here could be well fitted to 
describe ferromagnetism in these elements.

\vskip1.0cm 

\noindent {\Large {\bf 4. Conclusions}} 

\vskip0.5cm

In this paper we analyzed the influence on ferromagnetism of hopping interaction, $\Delta t$, and exchange-hopping interaction, $t_{ex} $. These interactions are nonzero when 
the hopping integral depends on the occupations of site involved in hopping.

The physical coupling between these interactions and the magnetic ordering comes from lowering the kinetic energy of the system during the transition to magnetic state by these 
interactions. They depend on the following two parameters: $S=t_1 /t_0 $ and $S_1 =t_2 /t_1 $.

The formalism, which was used in here, was the standard formalism for magnetism with correlation effects.\cite{4,11} The correlation effects lead to the change of shape of one 
spin band with respect to another and also to the bandwidth change of one spin band with respect to another. Mathematically they are described by the parameters $K_U $ and 
$K_b$.

The kinetic energy parameters ($\Delta t$ and $t_{ex} )$ influence the ferromagnetism also directly (in the spirit of Weiss theory) by contributing to the band shift factors the 
following quantities: $I_{\Delta t} $ and $I_{t_{ex}}$.

In effect we obtain numerous criteria for ferromagnetism, which show several physical features. 

The hopping interaction $\Delta t$ helps ferromagnetism for concentrations $n<1$ (see Fig. 1(a)).

The exchange-hopping interaction $t_{ex} $ helps ferromagnetism, but only weakly, for concentrations $n>1$ (see Fig. 2(a)).

The on-site Coulomb repulsion $U$ treated in the alloy analogy approximation helps the ferromagnetism mostly in the middle of the band (see Fig. 4).

Fig. 5 shows the critical on-site exchange interaction for ferromagnetism for the asymmetrical DOS with maximum density around the Fermi energy. Simple Stoner criterion depicts 
the strong tendency towards magnetism at the 
end of the band.

The next two figures display the results of critical on-site exchange interaction for the same DOS, when both kinetic energy parameters are present and are the function of $S=S_1 
$; $\Delta t=t_0 \left( {1-S}\right)$, and $t_{ex} =t_0 (1-S)^2/2$. One can see that when $t_2 <t_1 <t_0 $ ($S<1)$ the ferromagnetism is enabled at smaller concentrations (see Fig. 
6, where the curve with $S=0$ corresponds to the total exclusion of hopping in the presence of another 
electron). When $t_2 >t_1 >t_0 $ ($S>1)$ the ferromagnetism is enhanced quite dramatically at high concentrations (see Fig. 7). This situation can correspond to the ferromagnetic 
pure transition metals.

\end{document}